Title:     Spreading  of the conception  of permanent resonance  to wave motions over lattices

Authors: B.N.  Zakhariev. (Joit Institute for Nuclear Research, Dubna, Russian Federation,
e-mail : zakharev@theor.jinr.ru

Comments:  7  pages,  2 figures

Discovered by us [1] special (permanent) resonance mechanism of spectral zone creation in periodic structures is  generalized  to the case of  discrete space lattices and finite difference Schroedinger equation with local  V(n) and minimally nonlocal U(n). It would be interesting to generalize the periodicity formalism to the multidiagonal   finite-difference  potentials.

> This is as stupid as a fact, said Balzac. Thus he
> defined the concept of a fact in its own right. A fact
> in itself is of little interest until it is investigated  and
> interpreted and therefore becomes a phenomenon.
> Maxim Gorky

There are three main types of spectra: discrete, continuous , and band-like. The last consists of discrete pieces of continuous states, but the existence of spectral gaps between them  is physically  and mathematically a special phenomenon. Its  explanation by Bloch-Floquet formalism seemed to be perfect since its appearance about 100 years ago.  All this time there do not appeared an idea about improvement and simplification of the already achieved  notions. Maybe it was just due to the beauty of the original theory.  Besides this the simple physical cause of the tearing of the continuous spectrum  with creation of forbidden zones remained unmentioned.

We have discovered [1] that there  is strict coincidence of spatial oscillation frequencies of potential  and wave solutions. And this resonance together with some asymmetry of the wave bumps on each period destroyed  the physical solutions in continuous bands of energy spectrum. And what was for us  wonderful, that  this perfect  resonance appeared to be possible  not at separate energy points, but on full energy intervals of infinite (continuous) values of E, despite of the fact  that V(r) is **fixed**   and solutions are even non-periodic inside the forbidden interval and strongly shape dependent on energy. One of the important lessons here is the original universe approach to problem.  **Wave solutions** 'feel' the same potential quite differently due to distinguished  distributions of their most sensitive bumps and weakly  sensitive knots relative to spatial oscillations of potential (its partial barriers and wells).  The stronger is the overlap of the bumps with barriers, the more effectively repulsive  seems the influence of the fixed periodic potential. And the more the wave bumps embrace the partial wells of periodic perturbation, the stronger is its effective attraction. In intermediate case there is more attraction from the left or from the right  hand side. There is continuous change of 'average' strength of potential between two limit cases corresponding to the upper and lower boundaries of forbidden zone.  And this causes the asymmetry of wave perturbation on the periodic interval. The continuous variation of the effective averaged potential V  value on the interval   allows the complete compensation of total energy E variation  so that the effective kinetic energy (E- V) remains unchanged in some averaged sense. This  allows the existence of the exact permanent resonances over the whole forbidden zone. The key fact here is the periodicity only of  knots of special pairs of fundamental solutions. Inside each period the asymmetry in x of potential repeated hammering these solutions results in swinging waves (as the result of continuous joining waves of neighbor periods) with exponentially increasing (or decreasing) amplitudes in opposite directions. That makes such solutions un-normalizable and nonphysical. So are also all other solutions in forbidden zone because any solution is 'constructed' of un-normalizable fundamental constituents (see more detailed explanation in [1]. It was instructive to return to our old idea of creation of spectral gap at the chosen position of energy level (eigenstate) in infinite potential well with vertical walls at the ends of the period and the bottom shape exactly equal to the section of periodic potential on the period.

It was interesting to see whether the same intuitive interpretation of gap creation mechanism is valid in the case of finite difference Schroedinger equation [1,2]
.
$$- [\Psi(n+1) - 2\Psi(n) + \Psi(n-1)]/\Delta = \Delta [E - V(n)] \Psi(n) \qquad (1)$$

The difficulty here was that continuous relative space shift of potential and wave solution which was important in usual theory seems impossible in the case of discrete coordinate [ x-lattice]. Another difference of continuous and discrete space consist in that even for free motion of waves in discrete case there is only one allowed energy zone of finite width although there is no perturbation potential at all. It can be explained nevertheless that self the equidistant discrete coordinate values can serve for resonance destruction of physical solutions, - their swinging with exponentially increasing (decreasing) amplitudes

$$\exp(i\pi n) \exp(\kappa n) \; ; \; \exp(i\pi n) \exp(-\kappa n) \qquad (2)$$

In allowed zone the resonance conditions are violated and instead of exact infinite exponential growth of swinging amplitude there appears beatings. When the energy value shifts further from the zones boundary there appears violation of the resonance which increases with the distance $\Delta E$ from the forbidden zone. The difference in phases of potential and wave oscillations is also growing more intensive along the coordinate x-axis with $\Delta E$. At the places where these phases are near to one another the resonance is weakly violated and there is increase of wave swinging amplitude as in forbidden zone, but nevertheless after enough long x-interval the phases become so different that the regime of increasing to the right hand side of swinging amplitude changed to the opposite one and the amplitude begins to decrease. This is just the beatings in swinging which will be repeated more rapidly with the distance from forbidden zone [1].

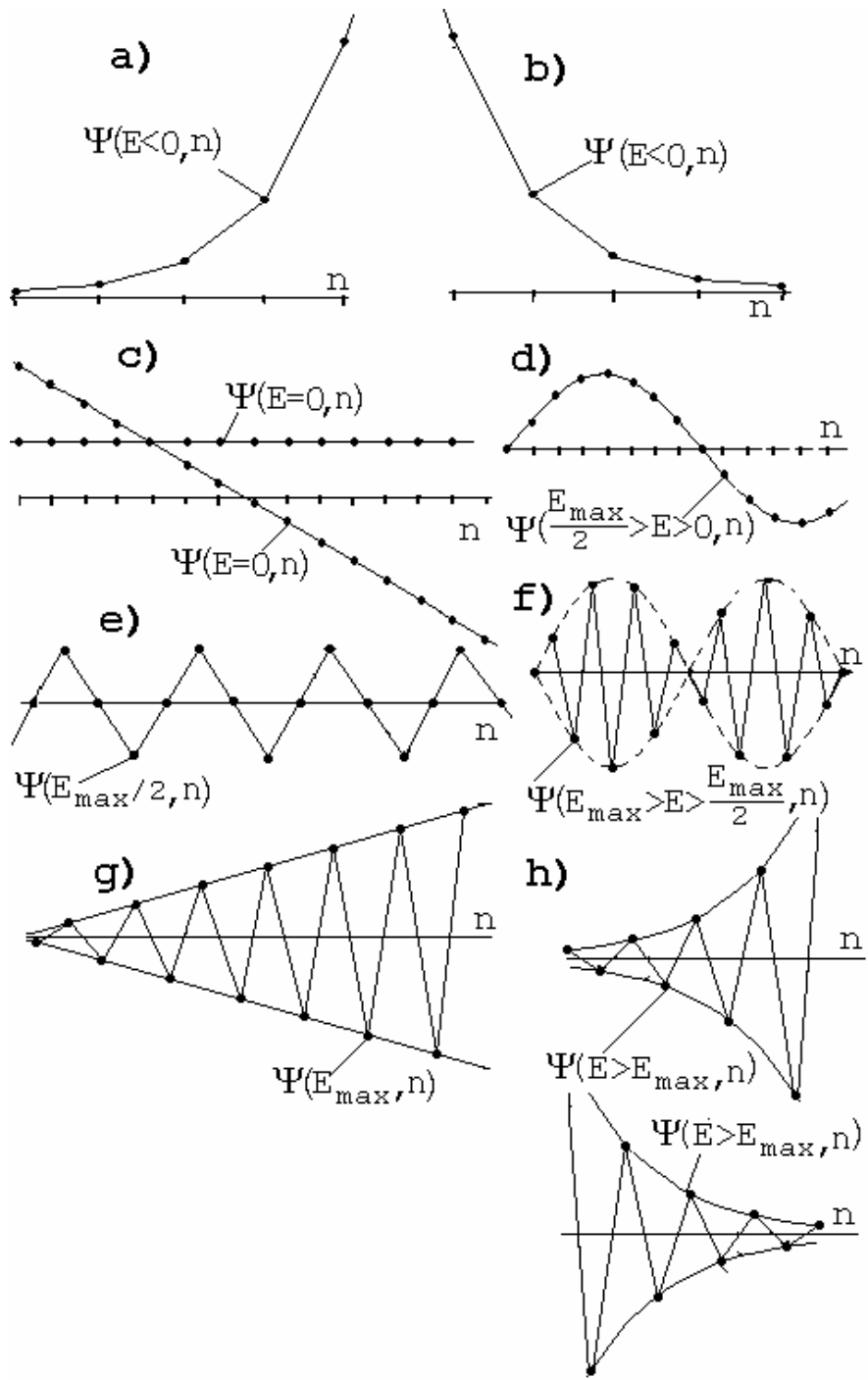

Fig.1. Solutions of the equation (1) with
V(x)=0 at different energies. Discrete values Ψ (n) are
connected for more obviousness by lines. When going through
the lower boundary of the allowed zone (E=0)  solution turns from
(a,b) exponential (at E<0) into (c) linear (E=Emin =0) and
then into (d,e,f) sine-like (E>0) function. The form of solution (e) in
the middle of allowed zone (E=Emax/2=2). The solution (f)
at energy below the upper boundary of the continuous spectrum. On
the upper boundary  E= Emax =4 the absolute values of the
solution (g) as also at E=$  is linear and for E>Emax
(h) is swinging with exponentially increasing amplitude what
corresponds to the forbidden zone as for E<0. Pay attention to
the symmetry of solutions  relative the middle of the allowed zone
up to the sign. The sine wave with 0<E< Emax /2 in Fig. d
corresponds to the sine-like enveloping line    (f) with energy on
interval  Emax /2 < E < Emax

The zig-zag fold sine is infinitely compressed with growth of energy in continuous case and in the finite difference case there is restriction by the lattice density : impossibility of oscillations which cannot be more frequent than discrete coordinate points.

As in the case of continuous space coordinate during transition from the forbidden to the allowed zone occurs the gradual violence of the resonance. So, near the zone boundary there appears swinging of solution almost as in forbidden zone. But the phase difference of wave and potential oscillations increased with x. And at some distance this caused the change of swinging regime : the swinging amplitude becomes decreasing instead of increasing. Further on x-axis there occurs again such change of swinging regimes. As a result we get beatings which a clearly shown in Fig.1 f. this is analogous to continuum case.

The stability of exact resonance conditions inside forbidden zones is due to the possibility of compensation of the variations of E by variations of
effective potential V in effective kinetic energy $(E-V(x))\Psi(x)$ dependent on relative positions of $\Psi$ bumps and V space oscillations .
The resonance nature of destroying of physical solutions appeared to be valid as in the case of free motion in outer zones and for the case of perturbation by periodic potentials.

If we switch on periodical perturbation on lattice there appears a forbidden zone inside the allowed zone of free motion over the lattice. In finite-difference Schroedinger equation the boundary conditions can be given in two neighbor points. And continuous changing these boundary conditions can be some analogy of continuous shifting coordinate x, although now we have only discrete coordinate points.

So, despite the difference in mechanisms of destruction of physical solutions on the continuous and lattice axis we succeeded in search of resonance essence of the phenomena unifying and enriching the general notion about it (spreading of quantum and mathematical intuition, see Also [2-9]).

We have considered the simplest form of the finite-difference Schroedinger equation (1) without nondiagonal elements of the interaction matrix potentials. But it would be interesting to generalize the consideration of wave solutions in periodic three-diagonal potentials $V(n)$ and $U(n)$. For simplicity we can at first restrict ourselves with the case with $V(n) = 0$. The minimally nonlocal potential $U(n)$ connects the values of wave function in neighbor points and is displaced on two nearest diagonals to the middle one and determines the local width of the allowed zone [2,3] . Unlike $U(n)$ the ordinary local potential $V(n)$, disposed on the main diagonal of the interaction matrix, and determines the bending of allowed zone without change its local , see Fig.8.22 and Fig. 8.21 in our new book (2008). To reduce the description of something unusual nonlocality to habitual local case here can be useful some analogy with multichannel problems where we discovered the inversion phenomenon : the possibility of attraction by potential barriers and repulsion by potential wells which was not previously mentioned exotic property of nondiagonal elements of interaction matrix [5] arXiv:0805.075 .

$$-[\Psi(n+1) - 2\Psi(n) + \Psi(n-1)]/\Delta^2 = [E - V(n)]\Psi(n) - U(n)\Psi(n+1) - U(n-1)\Psi(n-1) \quad (3)$$

Here the right hand side without E can be considered as an 'effective' local potential at the point n acting on $\Psi(n)$ at the same point n:

$$[V(n)+U(n)\Psi(n+1)/\Psi(n)+U(n-1)\Psi(n-1)/\Psi(n)]\Psi(n) \qquad (4)$$

And if $V(n)$ and $U(n)$ are periodic: $V(n+m)=V(n)$; $U(n+m)=U(n)$; $U(n-1+m)=U(n-1)$, it can be expected that the whole effective potential is also periodic with period m:

$$[V(n)+U(n)\Psi(n+1)/\Psi(n)+U(n-1)\Psi(n-1)/\Psi(n)]=$$
$$[V(n+m)+U(n+m)\Psi(n+1)/\Psi(n)+U(n-1)\Psi(n-1)/\Psi(n)] \qquad (5)$$

Here unexpectedly appeared a question which seemingly destroyed our plan to find the analogy of resonance mechanism of gap creation through periodic hammering the wave function and swinging its oscillations violating the normalization of solution making it unphysical as in the case of continuum coordinate. The point is that waves even in periodic potentials are periodic only in very special cases at the zones boundaries. And in our example wave functions enter inside the expression of the effective potential, what seems to ruin our hopes of explanation a new generalization of universal spectral gap creation mechanism. It was like a catastrophe. No hopes to find the exit in situation without way out and no sense to search it. .

And as it happens sometimes in scientific research, suddenly a simple hypothesis appeared. Here helped the memory of properties of some (most simple) fundamental solutions inside forbidden zones which were although not periodic, but with periodic knots and difference of wave values at any neighbor periods by the same factor. This was a key for the idea that in the fractions of neighbor values of functions must be almost all cancelled except the constant factors (this may be true for motion with constant $U(n)$. I have another period of hesitation abut the possibility to understand the more general case with periodic, not const $U(n)$. But as it happened not rarely, soon the fright of situation with no way out finished with a new understanding, that, although the values like $\Psi(n+1)/\Psi(n)$ and $\Psi(n-1)/\Psi(n)$ are not constant as for $U(n)=$const, but they are periodic due to cancellation of factor of amplitude growth for special fundamental solutions although separate such solutions are not periodic .What only violates the periodicity of the special fundamental functions in numerator and denominator is some constant. So, these factors will be cancelled in the fractions and the ratio of different periodic functions with the same period must be also periodic. It could be a reason to jump of joy and one more instructive example of usefulness not to hurry with sinking into despair during the research. Although not once the hope to reach the success was loosed, my brain continued to go over different ways of search the solutions being loaded enough with information, particularly about our original achievements in this area of research. So the problem with minimally nonlocal potential was reduced to the effective local periodic one, already considered above, see also the continuous case e.g., by Flugge [10].

It would be interesting to generalize the periodicity formalism to the multidiagonal finite-difference potentials. It seems that the procedure which we applied to three-diagonal potential for it reduction to one-diagonal effective potential must be valid also here.
  Does the more complicated boundary conditions change something important in different approaches.

It would be desirable to get more clear notion on how the properties of systems are transformed when the system becomes of finite size (because we have in our habitual practice to do with finite volumes). It seems probable that on macro distances the properties of finite system does not differ too much of infinite ones.